\documentclass[aps,a4paper,prd,twocolumn,showpacs,superscriptaddress,10pt,nofootinbib]{revtex4}
\usepackage{yfonts}
\usepackage{amssymb}
\usepackage{amsmath}

\usepackage[dvips]{epsfig}

\newcommand{\be}{\begin{equation}}
\newcommand{\ee}{\end{equation}}
\def\bea{\begin{eqnarray}}
\def\eea{\end{eqnarray}}

\newcommand{\cl}{{\cal{C}} \kern -0.1em \ell}     %Clifford algebra

\newcommand{\bn}{\begin{eqnarray}}
\newcommand{\en}{\end{eqnarray}}

\newcommand{\ii}{\textgoth{i}}
\newcommand{\jj}{\textgoth{j}}
\newcommand{\kk}{\textgoth{k}}
\newcommand{\CC}{\mathbb{C}}

\def\bea{\begin{eqnarray}}
\newcommand{\BK}{\mathbb{K}}

\def\eea{\end{eqnarray}}

\newcommand{\beq}{\begin{eqnarray}}
\newcommand{\eeq}{\end{eqnarray}}
\begin{document}

\title{Questing for Algebraic Mass Dimension One Spinor Fields}
\author{{C. H. Coronado Villalobos}}
\email{ccoronado@feg.unesp.br}
\author{{J. M. Hoff da Silva}}
\email{hoff@feg.unesp.br}
\affiliation{UNESP Universidade Estadual Paulista - Campus de Guaratinguet\'a - DFQ.\\
Av. Dr. Ariberto Pereira da Cunha, 333 CEP 12516-410, Guaratinguet\'a - SP,
Brazil.}
\author{{Roldao da Rocha}}
\email{roldao.rocha@ufabc.edu.br}
\affiliation{CMCC,
Universidade Federal do ABC, 09210-580, Santo Andr\'e, SP, Brazil}

\begin{abstract}
This work deals with new classes of spinors of mass dimension one in Minkowski spacetime. In order to accomplish it, the Lounesto classification scheme and the inversion theorem are going to be used. The algebraic framework shall be revisited by explicating the central point performed by the Fierz aggregate. Then the spinor classification is generalized in order to encompass the new mass dimension one spinors. The spinor operator is shown to play a prominent role to engender the new mass dimension one spinors, accordingly.
\end{abstract}

\maketitle

%\newpage
\section{Introduction}

There is a spinor classification  due to Lounesto \cite{lou}, which is particularly interesting for physicists due to its twofold ubiquitous aspect: on the one hand it is based upon bilinear covariants, and thus upon physical observables. On the other hand, by a peculiar multivector structure  --- the Fierz aggregate --- that leads to the so-called boomerang \cite{lou}, a quite elegant geometrical interpretation may be added to the classification. Moreover, with the aid of the boomerang it is possible likewise to prove that there are precisely six different classes of spinors in Lounesto's classification \cite{lou}. The most general form of the respective spinors in each class were introduced in  \cite{Cavalcanti:2014wia}. Lounesto's spinor classification was further employed to derive all the Lagrangians for gravity from the quadratic spinor Lagrangian  \cite{daRocha:2007sd}.
Higher dimensional spaces have a similar spinor classification \cite{bonora}, however the so-called geometric Fierz identities \cite{Lazaroiu:2013kja} obstruct the proliferation of new spinors classes in higher dimensions \cite{bonora}.

Within the Lounesto classification, a specific bilinear covariant plays a crucial role, since it can not be zero. This bilinear represents current density, at least  for the case of a regular spinor describing the electron. Its  components read  $J_\mu = J_\mu e^\mu = \psi^\dagger \gamma_0\gamma_\mu\psi e^\mu$, where $\psi$ denotes a spinor and $e^\mu$ is a dual basis in $\cal{C}\ell_{\mathrm{1,3}}$. Additionally, it is valuable to remark that ${\bf J}=J_\mu e^\mu$ is essential for the  definition of the boomerang structure. Regarding the electron theory, it is straightforward to realize the physical argument to explain why ${\bf J}$ must not vanish. Indeed, ${\bf J}$ is the conserved current in this case and therefore if ${\bf J}=0$ there is no associated particle \cite{bj}. In particular the time component $J_0 = \psi^\dagger\psi$ provides
the probability density of the electron, and when integrated over the spacetime it should be obviously non-null.

One of the main points that shall be pursued in this work is that ${\bf J}$ can be understood as a conserved current solely when the considered spinor obeys the usual dynamics rules by the Dirac equation, namely, it is an eigenspinor of the Dirac operator or, equivalently, it is described by the Dirac Lagrangian. The canonical mass dimension in this case is the same mass dimension $3/2$ associated to usual spin-1/2 fermions in the standard model. Since we are looking for possible manifestations of mass dimension one fermions in Minkowski spacetime, it is indeed possible to set ${\bf J}=0$, accordingly. In fact, by accomplishing it, even the previously mentioned algebraic argument precluding new spinor classes may be circumvented. Nevertheless, in this novel context, we should emphasize that the underlying dynamics shall not be dictated by the well-known Dirac equation. As the construction is relativistic,  the spinors arising from the analysis with ${\mathbf J}=0$ shall respect \emph{a priori} merely the Klein-Gordon equation. Actually, in a very conventional scheme, they must do so. Hence, the epigraph is now explained: the resulting spinors must have mass dimension one. Clearly by ``mass dimension'' we mean the canonical mass dimension of the associated quantum field, which inherits this property from the dynamics respected by its expansion coefficients.

Mass dimension one spinors have attracted attention mainly due to the fact that they can be coupled only to gravity and to scalar fields as well as, in a perturbatively renormalizable way. It thus makes it suitable for exploration under the ensign of dark matter. Mass dimension one spinors in Minkowski spacetime known in the literature are the so-called Elko spinors, which have been studied in a comprehensive context. They  comprise prominent applications in 4D gravity and cosmology \cite{daRocha:2007sd,Wei:2010ad,5307,daSilva:2014kfa,S.:2014dja,Agarwal:2014oaa},  and in brane-world models  as well \cite{Liu:2011nb,Jardim:2014xla}, besides their exotic counterparts \cite{daRocha:2011yr, Bernardini:2012sc}. Moreover, despite of the robust and rich
framework already developed \cite{Ahluwalia,Ahluwalia:2010zn,Ahluwalia:2008xi,Julio,Wunderle:2010yw}, Elko has been predicted to be measured in Higgs processes at LHC \cite{LHC,daSilva:2012wp} and explored in tunnelling methods concerning black holes \cite{hawk}. Massive spin-1/2 fields of mass-dimension were obtained  by constructing quantum fields from higher-spin Elkos, however these fields are still linked to the Elko construct.
 We stress, however, that the spinors to be found here are intrinsically different from the Elkos by the simple fact that ${\bf J}\neq 0$ in the Elko case.

The classification of mass dimension one spinors is performed by a possible and consistent modification in the Lounesto classification. However, in order to have an explicit form for them it is necessary the use of the so-called inversion theorem \cite{Cra, Taka}.

This paper is organized as follows: in the next Section the main steps of the framework which supports our analysis shall be revisited, namely the standard Lounesto classification and the inversion theorem. In Section III we show the existence of three new classes of mass dimension one spinors, obtaining the algebraic form in each case accordingly. In the last Section we point our concluding remarks and a pave a brief outlook.

\section{The Framework}

In order to properly address   the problem to be approached and solved, it is pivotal to review some key aspects of the standard formalism, highlighting the structures to be studied and generalized. To start,  the Lounesto's spinor classification shall be revisited, and subsequently  the inversion theorem algorithm shall be thereafter employed, accordingly.

\subsection{The Lounesto's Spinors Classification and  Generalizations}

Consider the Minkowski spacetime $(M,\eta_{\mu\nu})$ and its tangent bundle $
TM$. Denoting sections of the exterior bundle by
$\sec\Lambda (TM)$, given a $k$-vector $a \in \sec\Lambda^k(TM)$,  the {reversion} is defined by $\tilde{a}=(-1)^{|k/2|}a$, whilst the grade
involution reads $\hat{a}=(-1)^{k}a$, where $|k|$ stands for the integral part of $k$. By extending the Minkowski metric  from $\sec\Lambda^1(TM)=\sec T^*M$ to
$\sec\Lambda(TM)$, and  considering  $a_1,a_2 \in \sec \Lambda(V)$, the
{left contraction} is given by ${g}(a \lrcorner a_1,a_2)={g}(a_1
,\tilde{a}\wedge a_2 ). $
The well-known Clifford product between (the dual of) a vector field $ v \in \sec\Lambda^1(TM)$ and a multivector is prescribed  by $ v a = v \wedge a+ v
\lrcorner a $, defining thus the spacetime Clifford algebra $\cl_{1,3}$.  The set $\{{e}_{\mu }\}$ represents sections of the frame bundle
$\mathbf{P}_{\mathrm{SO}_{1,3}^{e}}(M)$ and  $\{\gamma^{\mu }\}$ can be further thought as being  the dual
basis $\{{e}_{\mu }\}$, namely, $\gamma^{\mu }({e}_{\mu })=\delta^\mu_{\;\nu}$.
 Classical spinors are objects of the space that carries the usual
$\tau=(1/2,0)\oplus (0,1/2)$ representation of the Lorentz group, that  can be thought as being sections of the vector bundle
$\mathbf{P}_{\mathrm{Spin}_{1,3}^{e}}(M)\times _{\tau }\mathbb{C}^{4}$.

Given a
spinor field
$\psi \in \sec \mathbf{P}_{\mathrm{Spin}_{1,3}^{e}}(M)\times_{\tau
}\mathbb{C}^{4}$, the bilinear covariants  are sections of the bundle $\Lambda(TM)$ \cite{lou,Cra}.
Indeed, the well-known Lounesto's spinors classification is based upon bilinear covariants and the underlying multivector structure. The physical nature of the classification focuses on the bilinear covariants, that are physical observables, characterizing types of fermionic particles. The observable quantities are given by the following multivector structure:
\begin{eqnarray}
\label{cova}
\sigma&=&\psi^{\dag}\gamma_{0}\psi, \hspace{1.9cm}  \omega=-\psi^{\dag}\gamma_{0}\gamma_{0123}\psi,\nonumber\\ J_\mu&=&\psi^{\dag}\mathrm{\gamma_{0}}\gamma_{\mu}\psi,  \hspace{1cm}
K_\mu=\psi^{\dag}\mathrm{\gamma_{0}}\textit{i}\mathrm{\gamma_{0123}}\gamma_{\mu}\psi,\nonumber\\ S_{\mu\nu}&=&\frac{1}{2}\psi^{\dag}\mathrm{\gamma_{0}}\textit{i}\gamma_{\mu\nu}\psi,
\end{eqnarray}
 where $\gamma_{0123}:=i\gamma_5=\gamma_0\gamma_1\gamma_2\gamma_3$. The set $\{\mathbf{1},\gamma
_{I}\}$ (where $I\in\{\mu, \mu\nu, \mu\nu\rho, {5}\}$ is a composed index) is a basis for
${\cal{M}}(4,\mathbb{C})$ satisfying  $\gamma_{\mu }\gamma _{\nu
}+\gamma _{\nu }\gamma_{\mu }=2\eta_{\mu \nu }\mathbf{1}$.

The above bilinear covariants in the Dirac theory are interpreted respectively  as the  mass of the particle ($\sigma$), the pseudo-scalar ($\omega$) relevant for parity-coupling, the current of probability ($\mathbf{J}$), the direction of the electron spin ($\mathbf{K}$), and the probability density of the intrinsic electromagnetic moment ($\mathbf{S}$) associated to the electron. The most important bilinear covariant for the our goal here  is $\mathbf{J}$, although with a different meaning. In fact, in the next section we shall set $\mathbf{J}=0$, enabling the extension of the standard Lounesto's classification to this case.

A prominent requirement for the Lounesto's spinors classification is that the bilinear covariants satisfy quadratic algebraic relations, namely, the so-called Fierz-Pauli-Kofink (FPK) identities, which read
\begin{eqnarray}
\label{Fierz}
J_{\mu}J^{\mu}&=&\sigma^{2}+\omega^{2},\;\;\;\;\; J_{\mu}J^{\mu}=-K_{\mu}K^{\mu},\nonumber\\
J_{\mu}K^{\mu}&=&0,\;\;\;\;\;
\mathbf{J}\wedge\mathbf{K}=-(\omega+\sigma\gamma_{0123})\mathbf{S}.
\end{eqnarray}
It is worth to remark that the above identities are fundamental, not merely for the aims regarding the classification, however for moreover asserting the inversion theorem, as we are going to see in the next subsection.

Within the Lounesto classification scheme, a non-vanishing $\mathbf{J}$ is crucial, since it enables to define the so-called boomerang \cite{lou} which has an ample geometrical meaning to assert that there are precisely six different classes of spinors. This is a prominent consequence of the  definition of a boomerang \cite{lou}. As far as the boomerang is concerned, it is not possible to exhibit more than six types of spinors, according to the bilinear covariants. Indeed, Lounesto's spinor classification splits regular and singular spinors. The regular spinors are those which have  at least one of the  bilinear covariants $\sigma$ and $\omega$ non-null. On the other hand, singular spinors present $\sigma=0=\omega$, and in this case the Fierz identities are in general replaced
by the more general conditions \cite{Cra}:
\begin{eqnarray}
Z^{2}=4\sigma Z,\qquad Z\gamma_{\mu}Z=4J_{\mu}Z,\qquad Z\gamma_{0123}Z=-4\omega Z\nonumber\\ Zi\gamma_{\mu\nu
}Z=4S_{\mu\nu}Z,\qquad Zi\gamma_{0123}\gamma_{\mu}Z =4K_{\mu}Z.
\end{eqnarray}

When an arbitrary spinor $\xi$ satisfies $\widetilde{\xi^{*}}\psi\neq 0$ and belongs to $\mathbb{C}\otimes\mathcal{C\ell}_{1, 3}$ --- or equivalently when $\xi^{\dagger}\gamma_{0}\psi\neq0\in {\cal{M}}(4,\mathbb{C})$ ---- it is possible to recover the original spinor $\psi$ from its aggregate $\mathbf{Z}$ given by
\begin{eqnarray}
\mathbf{Z}=\sigma + \mathbf{J} +i\mathbf{S}+i\mathbf{K}\gamma_{0123}+\omega \gamma_{0123}\, \label{Z}
\end{eqnarray}
and the spinor $\xi$ by the so-called Takahashi algorithm \cite{Taka} likewise. In fact, the spinor $\psi$
and the multivector ${\rm{\bf{Z}}}\xi$ differ solely by a multiplicative constant, and  can be thus
written as
\begin{equation}
\psi=\frac{1}{2\sqrt{\xi^{\dagger}\gamma_{0}\mathbf{Z}\xi}}\;e^{-i\theta}\mathbf{Z}\xi, \label{3}%
\end{equation}
\noindent where
$e^{-i\theta}={2}({\xi^{\dagger}\gamma_{0}\mathbf{Z}\xi})^{-1/2}\xi^{\dagger}\gamma_{0}\psi\in {\rm U(1)}$. For more details see,
e.g., \cite{Cra}. Equivalently to Eq.(\ref{3}),   we
shall use hereupon the notation $\psi\backsim\mathbf{Z}\xi$ to say that both sides of this equivalence are in the same equivalence class with respect to the quotient by $\mathbb{C}$. Moreover, when  $\sigma,\omega,\mathbf{J},\mathbf{S}%
,\mathbf{K}$ satisfy the Fierz identities, then the complex multivector
operator $\mathbf{Z}$ is named a {Fierz aggregate}. When $\gamma
_{0}\mathbf{Z}^{\dagger}\gamma_{0}=\mathbf{Z}$, thus $\mathbf{Z}$ is said to be a
 {boomerang} \cite{lou}.

The Takahashi algorithm reveals the importance of the aggregate. Moreover, the inversion theorem (to be regarded in the next subsection) is inspired on this spinor representation (\ref{3}). More significantly here, the aggregate plays a central role within the Lounesto classification since, in order to complete the classification itself, $\mathbf{Z}$ have to be promoted to a boomerang, satisfying
\begin{eqnarray}
\mathbf{Z}^{2}=4\sigma\mathbf{Z}. \label{boom}
\end{eqnarray} Obviously, for the regular spinors case the above condition is satisfied and $\mathbf{Z}$ is automatically a boomerang. However, for singular spinors case it is not so direct. Indeed, for singular spinors we must envisage the geometric structure underlying to the multivector. From the geometric point of view the following relations between the bilinear covariants must be fulfilled in order to ensure that the aggregate be a boomerang: $\mathbf{J}$ must be parallel to $\mathbf{K}$ and both are in the plane formed by the bivector $\mathbf{S}$. Hence, using the Eq. (\ref{Z}) and taking into account that we are dealing with singular spinors, it is straightforward to see that the aggregate can be recast as
\begin{eqnarray}
\mathbf{Z}=\mathbf{J}(1+i\mathbf{s}+ih\gamma_{0123}), \label{ZB}
\end{eqnarray} where $\mathbf{s}$ is a space-like vector orthogonal to $\mathbf{J}$, and $h$ is a real number. The multivector as expressed in Eq. (\ref{ZB}) is a boomerang \cite{Julio}. By inspecting the condition (\ref{boom}) we see that for singular spinors $\mathbf{Z}^2=0$. However, in order to the FPK identities hold it is also necessary that both conditions\footnote{We remark that $\mathbf{J}$ must be different from zero in the Lounesto classification.} $\mathbf{J}^2=0$ and $(\mathbf{s}+h\gamma_{0123})^2=-1$ must be satisfied. These considerations are important in order to constrain the possible spinor classes.

Now, let us explicit that from (\ref{3}) one can see that different bilinear covariants combinations may lead to different spinors, taking into account the constraints coming from the FPK identities. Altogether, the algebraic constraints reduce the possibilities to six different spinor classes, namely:
\begin{enumerate}
  \item $\sigma\neq0$, $\quad\omega\neq0$;
  \item $\sigma\neq0$, $\quad\omega=0$;
  \item $\sigma=0$, $\quad\omega\neq0$;
  \item $\sigma=0=\omega,$ \hspace{0.5cm} $\textbf{K}\neq0,$ $\quad\textbf{S}\neq0$;
  \item $\sigma=0=\omega,$ \hspace{0.5cm} $\textbf{K}=0,$ $\quad\textbf{S}\neq0$;
  \item $\sigma=0=\omega,$ \hspace{0.5cm} $\textbf{K}\neq0,$ $\quad\textbf{S}=0$.
\end{enumerate}

The spinors types-(1), (2) and (3), are called Dirac spinor fields (regular spinors). The spinor field (4) is called flag-dipole \cite{FRJR}, while the spinor field (5) is named flag-pole \cite{Benn}. Majorana \cite{Maj} and Elko \cite{Ahluwalia,Julio} spinors are elements of the flag-pole class. Finally, the type (6) dipole spinors are examplified by Weyl spinors. Note that there are only six different spinor fields. To see that, notice that for the regular case  since $\mathbf{J}\neq 0$ it follows that $\mathbf{S}\neq 0$ and $\mathbf{K}\neq 0$ as impositions from the identities (\ref{Fierz}). On the other hand, for the singular case, the geometry asserts that $\mathbf{J}(\mathbf{s}+h\gamma_{0123})=\mathbf{S}+\mathbf{K}\gamma_{0213}$. Hence, as far as $\mathbf{J}\neq 0$, as have already considered all the possibilities.

As it is clear from the above reasoning, $\mathbf{J}\neq 0$ is much more a matter of taste. It is instead an algebraic necessity on demonstrating the existence of six different class. In fact however a non vanishing $\mathbf{J}$ is indispensable only for the regular spinor case. As mentioned, the above classification makes use of this constraint in all the cases, since the very idea of the classification was to categorize spinors which could be related to Dirac particles in some aspect. As far as we leave this (physical) concept, more spinors can be found.

By taking $\mathbf{J}=0$, we cannot describe Dirac particles anymore. Therefore, the spinors arising from this consideration must  be merely ruled by the Klein-Gordon dynamics and, therefore, they must have mass dimension one. We finalize by stressing that the resulting spinors (see Section 3) have to be singular, as in contrary they would violate the FPK identities and, besides, the geometrical aspects underlying the algebraic structure need to be reconsidered.

%%%%%%%%%%%%%%%%%%%%%%%%%%%%%%%%%%%%%%%%%%%%%%%%%%%%%%%%%%%%%
\subsection{The Inversion Theorem}

It is well known, in the quantum mechanical context, that all the physical observables are represented by quadratic quantities of the wave function, for example the probability density. In the specific case of the Dirac particle, represented by a four-component spinor wave function $\psi$, we can write sixteen real quadratic forms, called bilinear covariants $\rho_{i}=\widetilde{\psi}\Gamma_{i}\psi$. The bilinear covariants are represented in the set of Eqs. (\ref{cova}). The bilinear covariants are not individual quantities \cite{Taka}, since their structure depends on the spinor itself. Crawford makes use of the FPK identities to define the  inversion theorem, which asserts that the general form of an arbitrary spinor may be expressed in terms of the bilinear covariants as
\begin{eqnarray}
\psi&=&\mathrm{e}^{-i\varphi}\Bigg(\Sigma-i\Pi\gamma_{5}+J_{\mu}\gamma^{\mu}-K_{\mu}\gamma_{5}\gamma^{\mu}+\frac{1}{2}S_{\mu\nu}\sigma^{\mu\nu}\Bigg)\xi,\nonumber\\
&=&\mathrm{e}^{-i\varphi}R^{i}\Gamma_{i}\xi,
\end{eqnarray}
where the set $\{\varphi, R^{i}\}$, contains real functions, and $\xi$ is an arbitrary constant spinor. It is clear that even if we choose a specific spinor $\xi$, we have the freedom to choose a set $\{\varphi, R^{i}\}$, since that the function $\psi$ contains only eight independent functions. Another important assertion, taken into account by Crawford is that the set of functions $R^{i}$ must always satisfy the corresponding equations from the FPK identities. A proof for this statement can be found in Ref. \cite{Cra}.

It is important to stress that the alluded inversion is not unique, since we can choose an arbitrary phase $\varphi$, and the constant spinor $\xi$. Thus, concerning the inversion program, it is fairly important to bear in mind that it is useful within the formal algebraic context. In the next section, we shall apply the inversion theorem in order to recover mass dimension one spinors coming from a suitable modification of the Lounesto's scheme.

%%%%%%%%%%%%%%%%%%%%%%%%%%%%%%%%%%%%%%%%%%%%%%%%%%%%%%%%%%%%%
\section{Algebraic construction of new spinors}
After briefly revisiting the equivalence among the
classical, algebraic, and operator spinor formulations in what follows, we
shall be able to analyse the possible constructions for the new mass dimension one spinors. Let us hence start by expressing an  arbitrary  multivector in ${\mathcal{C}}\ell _{1,3}$  as --- henceforth $e_{\mu}e_{\nu}e_{\lambda}=e_{\mu\nu\lambda}$:
 \begin{gather}
\Gamma = \alpha+\alpha^{\mu}{e}_{\mu}+\alpha^{\mu\nu}{e}_{\mu\nu}+\alpha^{\mu\nu\sigma}{e}_{\mu\nu\sigma}+\alpha^{0123}{e}_{0123}.
\label{spinor}
\end{gather}
Given the isomorphism ${\mathcal{C}}\ell _{1,3}\simeq {\mathcal{M}}(2,\mathbb{H})$, where hereupon $\mathbb{H}$ denotes the quaternionic ring,  a primitive idempotent
$f=\frac{1}{2}(1+{e}_{0})$ is taken to define a minimal left ideal ${\mathcal{C}}\ell _{1,3}f$. This is relevant, in particular,  to attain a spinor representation of ${\mathcal{C}}\ell _{1,3}$. The most general multivector  in ${\mathcal{C}}\ell _{1,3}f$ reads
\begin{eqnarray}
 \zeta =&(\beta^{1}+\beta^{2}{e}_{23}+\beta^{3}{e}_{31}+\beta^{4}{e}_{12})f+\nonumber\\
 &+(\beta^{5}+\beta^{6}{e}_{23}+\beta^{7}{e}_{31}+\beta^{8}{e}_{12}){e}_{0123}f.
\label{kkk}
\end{eqnarray}
\noindent
Since the identification  $\zeta = \Gamma f \in \cl_{1,3}f$ holds,  hence it implies the following equivalence between their respective components:
\begin{eqnarray}
\beta^{1}&=&\alpha+\alpha^{0},\hspace{0.3cm} \beta^{2}=\alpha^{23}+\alpha^{023},\hspace{0.2cm} \beta^{3}=-\alpha^{13}-\alpha^{013},\nonumber\\ \beta^{4}&=&\alpha^{12}+\alpha^{012},\hspace{0.2cm} \beta^{5} = -\alpha^{123}+\alpha^{0123},\hspace{0.2cm} \beta^{6}=\alpha^{1}-\alpha^{01},\nonumber\\ \beta^{7}&=&\alpha^{2}-\alpha^{02},\hspace{0.2cm}  \beta^{8}=\alpha^{3}-\alpha^{03}.
\end{eqnarray}%
\noindent
By denoting $\textgoth{i}={e}_{2}e_{3},\;\textgoth{j}={e}_{3}e_{1},$ and $\textgoth{k}={e}_{1}e_{2}$, it is clear that the set  $\{1, \textgoth{i},\textgoth{j},\textgoth{k}\}$ is a basis for the quaternion algebra~$\mathbb{H}$.  The two quaternions appearing as coefficients in~(\ref{kkk}), namely,
\begin{eqnarray}
q_1&=& \beta^1+\beta^2{e}_{23}+\beta^3{e}_{31}+\beta^4{e}_{12},\nonumber \\  q_2&=& \beta^5+\beta^6{e}_{23}+\beta^7{e}_{31}+\beta^8{e}_{12} \in \mathbb{H}\,,
\end{eqnarray} where $\mathbb{H}=f\cl_{1,3}f = {\rm span}_\mathbb{R}\{1, {e}_{23},{e}_{31},{e}_{12}\}$  commutes with $f$ and ${e}_{0123}$. Hence it yields the equality
$
q_1f + q_2{e}_{0123}f = fq_1 + {e}_{0123}fq_2,
$
 evincing that the left ideal $\cl_{1,3}f$ is in fact a right module over $\BK$ with a basis $\{f,{e}_{0123}f\}$.
Moreover, the orthonormal basis $\{e_{\mu }\}$ have an immediate standard representation
\begin{eqnarray}
{e}_{0}&=&%
{\scriptstyle\begin{pmatrix}
1 & 0 \\
0 & -1%
\end{pmatrix}}%
,\quad{e}_{1}=%
{\scriptstyle\begin{pmatrix}
0 & \textgoth{i} \\
\textgoth{i} & 0%
\end{pmatrix}}%
,\quad{e}_{2}=%
{\scriptstyle\begin{pmatrix}
0 & \textgoth{j} \\
\textgoth{j} & 0%
\end{pmatrix}}%
,\quad{e}_{3}=%
{\scriptstyle\begin{pmatrix}
0 & \textgoth{k} \\
\textgoth{k} & 0%
\end{pmatrix}}\,,\nonumber
\end{eqnarray}
\noindent
which consequently induce representations for the idempotent  $f$ and the multivector ${e}_{0123}f$:
$$
[f]={\scriptstyle\begin{pmatrix}
1 & 0 \\
0 & 0%
\end{pmatrix}} \quad \mbox{and} \quad
[{e}_{0123}f]=%
{\scriptstyle\begin{pmatrix}
0 & 0 \\
1 & 0%
\end{pmatrix}}.
$$
Therefore, a general element ${\Gamma}\in{\mathcal{C}}\ell_{1,3}$ can be expressed as 
\begin{eqnarray}  \label{quat}
\begin{pmatrix}
q_1&q_2\\q_3&q_4\end{pmatrix}
\in {\mathcal{M}}(2,\mathbb{H})
\end{eqnarray}
where $q_1 = \alpha + \alpha^0 + (\alpha^{23} + \alpha^{023})\textgoth{i} 
-(\alpha^{13} + \alpha^{013})\textgoth{j} + (\alpha^{12} + \alpha^{012})\textgoth{k}$, $q_2 = 
(\alpha^{0123}-\alpha^{123}) + (\alpha^1 - \alpha^{01})\textgoth{i} 
+(\alpha^2 - \alpha^{02})\textgoth{j} + (\alpha^3 - \alpha^{03})\textgoth{k}$, $q_3 = 
-(\alpha^{123} + \alpha^{0123}) + (\alpha^1 + \alpha^{01})\textgoth{i} 
+(\alpha^2 + \alpha^{02})\textgoth{j} + (\alpha^3 + \alpha^{03})\textgoth{k}$ and $q_4 = 
(\alpha - \alpha^0) + (\alpha^{23} - \alpha^{023})\textgoth{i} 
+(\alpha^{013}-\alpha^{13})\textgoth{j} + (\alpha^{12} - \alpha^{012})\textgoth{k}%
$

 A multivector $\Psi$ in the even subalgebra $\cl_{1,3}^{+}$ is named  {spinor operator}, reading
\begin{equation}
\Psi=\alpha+\alpha^{\mu\nu}e_{\mu\nu}+\alpha^{0123}e_{0123}\,.
\label{400}
\end{equation}%
\noindent From the point of view of Eq.~(\ref{quat}) it yields
\begin{eqnarray}
&&[\Psi] = {\scriptstyle\begin{pmatrix}
q_{1} & -q_{2} \\
q_{2} & q_{1}%
\end{pmatrix}}\nonumber\\
&&=
\begin{pmatrix}
\alpha+\alpha^{23}\textgoth{i}-\alpha^{13}\textgoth{j}+\alpha^{12}\textgoth{k} &
-\alpha^{0123}\!+\!\alpha^{01}\textgoth{i}\!+\!\alpha^{02}\textgoth{j}\!+\!\alpha^{03}\textgoth{k} \\\!\!\!
\alpha^{0123}\!-\!\alpha^{01}\textgoth{i}\!-\!\alpha^{02}\textgoth{j}\!-\!\alpha^{03}\textgoth{k} &
\alpha\!+\!\alpha^{23}\textgoth{i}\!-\!\alpha^{13}\textgoth{j}\!+\!\alpha^{12}\textgoth{k}
\end{pmatrix} .\nonumber
\end{eqnarray}
The isomorphisms $ {\mathcal{C}}\ell _{1,3}\frac{1}{2}(1+e_{0})\simeq {\mathcal{C}}\ell _{1,3}^{+}\simeq \mathbb{H}^{2} \simeq \mathbb{C}^{4}$ among vector spaces respectively evince the
correspondence among the algebraic,  the operatorial, and  the classical definitions of a spinor in Minkowski spacetime. Indeed, the spinor space $\mathbb{H}^{2}$ carries the
${(1/2,0)}\oplus {(0,1/2)}$  (or ${(1/2,0)}$ or ${(0,1/2)}$)  representations of the Lorentz group, and is isomorphic both to the minimal left ideal
${\mathcal{C}}\ell _{1,3}\frac{1}{2}(1+e_{0})$, that is equivalent to  the algebraic spinor, and to the even subalgebra ${\mathcal{C}}\ell _{1,3}^{+}$ that corresponds to the space of spinor operators \cite{lou3,conformal}. Thus  the Dirac spinor is expressed equivalently as:
\begin{eqnarray}
&&{\scriptstyle\begin{pmatrix}
q_{1} & -q_{2} \\
q_{2} & q_{1}%
\end{pmatrix}} [f]
%=
%\begin{pmatrix} q_1 & -q_2\\q_2 & \phantom{-}q_1 \end{pmatrix} \begin{pmatrix} 1 & 0\\0 & 0 \end{pmatrix}
= {\scriptstyle\begin{pmatrix} q_1 & 0\\q_2 & 0 \end{pmatrix}}\cong{\scriptstyle\begin{pmatrix} q_1\\q_2 \end{pmatrix}}\nonumber\\ &&=\left(
\begin{array}{c}
\alpha+\alpha^{23}\textgoth{i}-\alpha^{13}\textgoth{j}+\alpha^{12}\textgoth{k} \\
\alpha^{0123}-\alpha^{01}\textgoth{i}-\alpha^{02}\textgoth{j}-\alpha^{03}\textgoth{k}%
\end{array}%
\!\right) \!\in \cl_{1,3}f\simeq\mathbb{H}^2 \label{hh}
\end{eqnarray}%

\noindent
Now by employing the usual representation
\begin{eqnarray}
1   &\mapsto& {\scriptstyle\begin{pmatrix}  1 & 0\\ 0 & 1\end{pmatrix}}, \quad\ii \mapsto {\scriptstyle\begin{pmatrix}  i & \phantom{-}0\\ 0 & -i \end{pmatrix}}, \quad
\jj \mapsto {\scriptstyle\begin{pmatrix}  \phantom{-}0 & 1\\ -1 & 0 \end{pmatrix}}, \quad
\kk \mapsto {\scriptstyle\begin{pmatrix}  0 & i\\ i & 0 \end{pmatrix}}\,,
\nonumber
\end{eqnarray}
in 2 $\times 2$ complex matrices, the  spinor operator $\Psi$ in~(\ref{400})  can be viewed furthermore as a $4\times 4$ matrix, as follows:
\begin{eqnarray}
&&\begin{pmatrix}
\alpha\!+\!\alpha^{23}i & \!-\!\alpha^{13}\!+\!\alpha^{12}i & \!-\!\alpha^{0123}\!+\!\alpha^{01}i & \alpha^{02}\!+\!\alpha^{03}i \\
\alpha^{13}\!+\!\alpha^{12}i & c\!-\!\alpha^{23}i & \!-\!\alpha^{02}\!+\!\alpha^{03}i & \!-\!\alpha^{0123}\!-\!\alpha^{01}i\\
%%%%%%%%%%%%%%%%%%%%%%%%%%%%%%%%%%%%%
\alpha^{0123}\!-\!\alpha^{01}i & \!-\!\alpha^{02}\!-\!\alpha^{03}i & \alpha\!+\!\alpha^{23}i & \!-\!\alpha^{13}\!+\!\alpha^{12}i\\
\alpha^{02}\!-\!\alpha^{03}i & \alpha^{0123}\!+\!\alpha^{01}i & \alpha^{13}\!+\!\alpha^{12}i & \alpha\!-\!\alpha^{23}i\\
\end{pmatrix}\nonumber\\
&&\equiv
\begin{pmatrix}
\psi_1 & -\psi_2^{*} & -\psi_3 &  \phantom{-}\psi_4^{*}\\
\psi_2 & \phantom{-}\psi_1^{*} & -\psi_4 & -\psi_3^{*}\\
\psi_3 & -\psi_4^{*} & \phantom{-}\psi_1 & -\psi_2^{*}\\
\psi_4 & \phantom{-}\psi_3^{*} & \phantom{-}\psi_2 & \phantom{-}\psi_1^{*}
\end{pmatrix}.
\end{eqnarray}
The spinor $\psi $ lives in the left (minimal) ideal $(\mathbb{C}\otimes \cl _{1,3})f$, where    $ f=\frac{1}{4}(1+e_{0})(1+ie_{12})$ is an idempotent that equals ${\rm diag}(1,0,0,0)$ in the Dirac representation, making ${e}_{\mu }\mapsto\gamma _{\mu }\in {\cal{M}}(4,\mathbb{C}$). Hence it follows that
\begin{eqnarray*}
\psi \simeq
\begin{pmatrix}
\psi_1 & 0 & 0 & 0 \\
\psi_2 & 0 & 0 & 0 \\
\psi_3 & 0 & 0 & 0 \\
\psi_4 & 0 & 0 & 0%
\end{pmatrix}\in (\CC\otimes \cl _{1,3})f, \hspace{0.5cm} \mbox{or}\hspace{0.5cm}
\begin{pmatrix}
\psi_1 \\
\psi_2 \\
\psi_3 \\
\psi_4%
\end{pmatrix}\in \mathbb{C}^{4}\,,  \label{dire}
\end{eqnarray*}%
illustrating the usual prescription between the multivector $\psi$ and the classical Dirac spinor field.

In this context, the posed conundrum is thus reduced to the calculation of the spinor operator (\ref{400}), finding $\psi$ \cite{lou,Vaz}. Prior to  accomplishing it, however, it is necessary to define the bilinear covariants in terms of the spinor operator $\Psi$ \cite{lou3}:
\begin{eqnarray}
\sigma&=&\langle\Psi\widetilde{\Psi}\rangle_{0}, \;\;\;\;\;\; \omega=-\langle\Psi e_{5}\widetilde{\Psi}\rangle_{0}, \hspace{0.5cm}
J=\Psi e_{0}\widetilde{\Psi},\nonumber\\
\label{13}
S&=&\Psi e_{1}e_{2}\widetilde{\Psi}, \hspace{0.5cm} K=\Psi e_{3}\widetilde{\Psi},
\end{eqnarray}
where $e_{5}=e_{0}e_{1}e_{2}e_{3}$ and $\langle\;\,\cdot\;\,\rangle_{0}$ denotes the scalar part of the multivector taken into account.

It is important to highlight that the bilinear covariants in (\ref{cova}) provide 16 independent quantities. On the other hand, it is also possible to express the spinor as a function of such bilinear covariants with an arbitrary phase (see section 2.2), according to the Takahashi theorem \cite{Taka}. Thus, keeping in mind that the spinor exhibits only 8 degrees of freedom and the bilinear covariants have 16 degrees of freedom, it is necessary to use the Fierz identities. Such identities reduce the degrees of freedom to 7, being the extra degree of freedom associated to a phase factor\footnote{For completeness, by considering Pauli spinors we have 4 degrees of freedom whilst the Fierz identities give account of 3 of them. Again, the extra degree of freedom is associated to a phase \cite{Vaz}.}. Taking into account Eq. (\ref{hh}), it is usual, in order to reduce the degrees of freedom of $\Psi$, to define the following relation
\begin{eqnarray*}
\alpha\exp({e_{12}\theta})&\cong&\frac{1}{4}\bigg(\Psi\!+\!e_{0}\Psi e_{0}\!+\!e_{21}\Psi e_{12}\label{5}
\!+\!e_{210}\Psi e_{012}\bigg),
\end{eqnarray*}
where $\alpha$ is a constant and $\theta$ is an arbitrary phase. To find the constant $\alpha$, we use the complex conjugate of Eq. (\ref{5}), that for the algebra here considered is equivalent to the reversion. It yields the following expression:
\begin{eqnarray}
\alpha\exp({e_{21}\theta})\!\cong\!\frac{1}{4}\bigg(\widetilde{\Psi}\!+\!e_{0}\widetilde{\Psi} e_{0}+e_{12}\widetilde{\Psi} e_{21}\label{5,1}
+e_{012}\widetilde{\Psi} e_{210}\bigg),
\end{eqnarray}
and by multiplying Eqs. (\ref{5}) and (\ref{5,1}) we obtain
\begin{eqnarray}
\alpha^{2}=&&\frac{1}{16}\bigg(\sigma+e_{5}\omega+\mathbf{J}e_{0}+\mathbf{S}e_{21}-e_{0123}\mathbf{K}e_{210}+\mathbf{J}e_{0}+\sigma\nonumber\\&&+e_{5}\omega-e_{0}e_{0123}\mathbf{K}e_{21}+e_{0}\mathbf{S}e_{210}-e_{21}(\sigma+e_{5}\omega)e_{21}\nonumber\\
&&+e_{21}\mathbf{S}-e_{21}e_{0123}\mathbf{K}e_{0}-e_{21}\mathbf{J}e_{210}-e_{210}e_{0123}\mathbf{K}\nonumber\\\nonumber
&&+e_{210}\mathbf{S}e_{0}-e_{210}\mathbf{J}e_{21}-e_{210}(\sigma+e_{5}\omega)e_{210}\bigg).
\label{6}\end{eqnarray}
Making use of $e_\mu e_\nu
+ e_\nu e_\mu =2\eta_{\mu\nu}$, it yields \begin{eqnarray}
\label{12}
\alpha=\frac{1}{2}\bigg(\sigma+e_{5}\omega+\mathbf{J}e_{0}-\mathbf{K}e_{3}-\mathbf{S}e_{12}\bigg)^{1/2}.
\end{eqnarray} The final step to determine $\Psi$ in terms of $\alpha$ and its bilinear covariants is to multiply Eq. (\ref{5,1}), from which we get
\begin{eqnarray}
\Psi\alpha\exp({e_{21}\theta})\,\cong&&\frac{1}{4}\bigg(\Psi\widetilde{\Psi}+\Psi e_{0}\widetilde{\Psi} e_{0}+\Psi e_{12}\widetilde{\Psi} e_{21}\nonumber\\\label{7,2}
&&+\Psi e_{012}\widetilde{\Psi} e_{210}\bigg).
\end{eqnarray}
By using the relations (\ref{13}), the expression for $\Psi$ is given by
\begin{eqnarray*}
\label{13,1}
\Psi=\frac{1}{4\alpha}\bigg(\sigma+e_{5}\omega+\mathbf{J}e_{0}-\mathbf{K}e_{3}-\mathbf{S}e_{12}\bigg)\exp({e_{12}\theta}).
\end{eqnarray*}

Through Eq. (\ref{400}), it is possible to define the algebraic spinor $\psi$ by
\begin{eqnarray}
\label{14}
\psi\!\!=\!\frac{1}{4\alpha}\!\bigg(\!\!\sigma+e_{5}\omega\!+\!\mathbf{J}e_{0}\!-\!\mathbf{K}e_{3}\!-\!\mathbf{S}e_{12}\!\bigg)\!\exp({e_{12}\theta})\!\left(
\begin{array}{c}
1\\
0\\
0\\
0\\
\end{array}
\right).
\end{eqnarray}

By means of Eq. (\ref{14}) it is possible to recover the algebraic spinor from its bilinear covariants via the inversion theorem setup. Having completed the above program for the general case, the application to new mass dimension one spinors follows straightforwardly.

As remarked in Section $2$, the Lounesto classification is based upon the FPK identities. As far as these relations are satisfied, novel  possibilities involving spinors can be considered. We propose a classification of new spinors, arising from considering that the bilinear covariant $\mathbf{J}$ is always null and the aggregate associated ($\mathbf{Z}$) is  no longer a boomerang as well. On the other hand, the bilinear covariants still satisfy the identities (\ref{Fierz}). As emphasized by the previous analysis, this last requirement is important, since that we shall express the new algebraic spinors functional form.

The consideration that the bilinear covariants must satisfy the FPK identities with $\mathbf{J}=0$ reveals the existence of three new spinors. We shall finalize this section by evincing their bilinears and their algebraic structure.

   {\bf Case 1:} $\sigma=0=\omega$, $\mathbf{J}=0$, $\mathbf{K}\neq0$ and $\mathbf{S}\neq0$. It can be verified that all the FPK identities (\ref{Fierz}) are satisfied. Moreover, the aggregate (not a boomerang) associated for this spinor reads
  \begin{eqnarray}
  \mathbf{Z}=i(\mathbf{S}+\mathbf{K}e_{0123}),
    \end{eqnarray}
Finally, considering this particular arrangement of the bilinear covariants, the spinor operator  is given by
\begin{eqnarray*}
  \label{15}
\Psi\cong\frac{1}{2\sqrt{-K_{3}-S_{21}}}(-K e_{3}-S e_{21})\exp({e_{12}\theta}),
\end{eqnarray*} and the algebraic spinor amounts out to be
\begin{eqnarray*}
\psi=\frac{1}{2\sqrt{-K_{3}-S_{21}}}(-Ke_{3}-S e_{21})\exp({e_{12}\theta})\left(
\begin{array}{c}
1\\
0\\
0\\
0
\end{array}
\right).
\end{eqnarray*}
The next cases follow in straightforward  analogy:

   {\bf Case 2:} $\sigma=0=\omega$, $\mathbf{J}=0$, $\mathbf{K}=0$ and $\mathbf{S}\neq0$. Here, the FPK identities are also satisfied and the aggregate associated is simply given by
\begin{eqnarray}
\mathbf{Z}=i\mathbf{S}.
\end{eqnarray}
The spinor operator reads
\begin{eqnarray*}
\label{16}
\Psi\cong\frac{1}{2\sqrt{-S_{21}}}(-S e_{21})\exp({e_{12}\theta}),
\end{eqnarray*}
and the algebraic spinor can be written as
\begin{eqnarray*}
\psi=\frac{1}{2\sqrt{-S_{21}}}(-S e_{21})\exp({e_{12}\theta})\left(
\begin{array}{c}
1\\
0\\
0\\
0
\end{array}
\right).
\end{eqnarray*}

   {\bf Case 3:} $\sigma=0=\omega$, $\mathbf{J}=0$, $\mathbf{K}\neq0$ and $\mathbf{S}=0$, again the FPK identities hold, and the associated spinor operator has the following form:
\begin{eqnarray*}
\Psi\cong\frac{1}{2\sqrt{-K_{3}}}(-K e_{3})\exp({e_{12}\theta}),
\end{eqnarray*}
leading to the following algebraic spinor
\begin{eqnarray*}
\psi=\frac{1}{\sqrt{-K_{3}}}(-Ke_{3})\exp({e_{12}\theta})\left(
\begin{array}{c}
1\\
0\\
0\\
0
\end{array}
\right).
\end{eqnarray*}
The cases we have shown demonstrate the existence of three new classes of spinors not catalogued previously, which in particular, present mass dimension one in Minkowski spacetime. These spinors have the specific bilinear covariant ${\bf J}$ equal to zero. Since for spinors respecting the Dirac dynamics ${\bf J}$ is the conserved current, here we must be dealing with spinors obeying only the Klein-Gordon equation. Notice that it is a natural consequence, since a given spinor in this context is nothing but a section of the bundle comprised by SL$(2,\mathbb{C})$ and $\mathbb{C}^4$. Thus, it must respect a relativistic dynamics. From the mathematical point of view, instead, ${\bf J}\neq 0$ is also a necessary condition to promote the Fierz aggregate to a more meaningful quantity (in the geometrical context), the boomerang which, in turn, is essential in reducing the number of different spinor class to six in the Lounesto classification. In the consideration of ${\bf J}=0$ the classification itself is rebuilt and new spinors arise.

%%%%%%%%%%%%%%%%%%%%%%%%%%%%%%%%%%%%%%%%%%%%%
\section{Concluding Remarks and Outlook}

We have shown the existence of three new spinors of mass dimension one,  via the inversion theorem and a consistent modification of the Lounesto spinor field classification. This has been achieved considering the specific bilinear covariant ${\bf J}$ equal to zero. Physically, it means that the new spinors can not respect the Dirac dynamics, only the Klein-Gordon one, enabling thus the canonical mass dimension equal to one.

A word of caution may be added to these final remarks. As remarked along the text, the adopted procedure is consistent and bearing in mind the precedent opened by previous mass dimension one spinors (the Elkos), the spinors found may have several physical relevant aspects to be explored \cite{LHC}. This is, in fact, our belief concerning the generalization presented here. However, one must take into account that the classification and the algebraic functional form do not say much about the emergence of these spinors in nature. As it is, the quantities described in the cases 1, 2, and 3 of last section are mathematically well defined structures whose associated physical field would have interesting properties. The possibility of physical manifestation of such spinors are currently under investigation.

%, which, lead us to conclude, they are not associated with Dirac spinors, since, the bilinear covariant $J$, for Dirac spinors, can be interpreted as the a current and here the current is null.\\
%On the other hand, the existence of these spinors, only exists by a algebraic construction, therefore, this work leaves an open window to search for a physical interpretation of these object.

%%%%%%%%%%%%%%%%%%%%%%%%%%%%%%%%%%%%%%%%%%%%%%%%%%%%
\subsection*{Acknowledgements}

CHCV thanks to CAPES (PEC-PG) for the financial support, JMHS thanks to CNPq
(308623/2012-6; 445385/2014-6) for partial support. RdR is grateful to CNPq grants  473326/2013-2 and 303027/2012-6. RdR is grateful to Prof. Loriano Bonora, for the fruitful discussions.

\end{document}